%% file: main.tex
\documentclass[lettersize,journal]{IEEEtran}
\usepackage[T1]{fontenc}
\usepackage{array}
\usepackage{amsmath}
\usepackage{booktabs}
\usepackage{graphicx}
\usepackage{longtable}
\usepackage[hidelinks]{hyperref}
\usepackage{multirow}
\usepackage{ragged2e}
\usepackage{tabularx}
\usepackage{stfloats}
\usepackage{tablefootnote}
\newcolumntype{P}[1]{>{\RaggedRight\arraybackslash}p{#1}}
\newcolumntype{Y}{>{\RaggedRight\arraybackslash}X}

\begin{document}

\title{Toward Explainability in Multimodal Recommendation}

\author{
    Chen Liang$^{1}$,
    Jinghui Li$^{1}$,
    Xiuze Zhou,
    Zongyue Wang$^{*}$,
    Yuanguo Lin$^{*}$,
    Han Yu


    \IEEEcompsocitemizethanks{
        \IEEEcompsocthanksitem C. Liang, J. Li, Z. Wang and Y. Lin are with the School of Computer Engineering, Jimei University, China. Email: liangchen@jmu.edu.cn, jhli@jmu.edu.cn, wangzongyue@jmu.edu.cn and xdlyg@jmu.edu.cn.
        \IEEEcompsocthanksitem H. Yu is with the College of Computing and Data Science, Nanyang Technological University, Singapore. Email: han.yu@ntu.edu.sg.
        \IEEEcompsocthanksitem X. Zhou is with The Hong Kong University of Science and Technology (Guangzhou), Guangzhou, China. Email: xz.zhou@connect.hkust-gz.edu.cn.
        \IEEEcompsocthanksitem $^{*}$ Corresponding author
        \IEEEcompsocthanksitem $^{1}$ Co-first authors
    }

}

\maketitle

\begin{abstract}
Multimodal recommender systems effectively alleviate data sparsity and cold-start problems by integrating heterogeneous information, yet complex cross-modal representations and model decision-making processes also introduce new interpretability challenges. Despite the rapid growth of research on the interpretability of multimodal recommendation in recent years, relevant methods and evaluation criteria remain relatively fragmented, lacking a unified conceptual framework. To address this gap, this paper categorizes existing methods into modality explanation and post-hoc explanation based on the timing and manner of explanation generation, and further constructs a progressive system for each category. Building on this foundation, this paper synthesizes four categories of evaluation metrics from both objective quantitative and subjective perceptual perspectives, and reviews relevant datasets along the dimensions of application domain and data source. Through a comprehensive analysis of existing research, we further propose future research directions. This paper aims to establish a unified analytical perspective across multimodal representation, recommendation decision-making, and user preferences, providing a research foundation for building transparent and trustworthy multimodal recommender systems.
\end{abstract}


\section{Introduction}

\IEEEPARstart{R}{ecommender} systems have become a core function of platforms such as e-commerce, social media, and e-learning \cite{lin2023survey}. Traditional recommender systems primarily rely on user behavior analysis data, generating recommendations by modeling user preferences and item attributes; however, they face challenges such as data sparsity and cold start. To address this issue, multimodal recommender systems \cite{xu2026survey} have been introduced. Multimodal data such as images, text, and audio contain rich semantic information and can complement user interest modeling from multiple dimensions. Multimodal data can provide richer user preference information than a single modality, thereby improving recommendation accuracy.

By fusing heterogeneous modality features, multimodal recommender systems break through the information bias bottleneck of traditional recommender systems. Applied research on short-video platforms demonstrates that fusing multimodal information such as video, audio, and text can significantly improve the accuracy of user interest modeling \cite{zhao2025multimodal}. Research in the fashion recommendation domain further validates the effectiveness of multimodal fusion, achieving more accurate recommendations by integrating product images and textual descriptions \cite{wu2020visual}. In e-commerce recommendation scenarios, it has been found that multimodal feature fusion can effectively alleviate the data sparsity problem \cite{laenen2022learning}.

Although multimodal recommendation demonstrates great potential, its ``black-box'' problem constrains the practical application of such systems. Users not only need to know ``what is recommended'' but also desire to understand ``why it is recommended.'' Interpretability has become a key challenge in the practical application of recommender systems \cite{guo2023towards}. In high-risk decision-making scenarios such as healthcare and finance, user demand for interpretability has reached an unprecedented level \cite{liang2025enhancing}. How to achieve interpretability within the multimodal recommendation framework has become a research hotspot attracting joint attention from both academia and industry \cite{alamdari2025edgex}.

\subsection{The Need for Explainability}

The interpretability of a recommender system refers to the system's ability to clarify the basis of recommendation decisions for users. Recommender systems with interpretability can significantly improve user trust and satisfaction \cite{guo2023towards}. The positive impact of interpretability on user acceptance has been verified in conversational recommendation scenarios \cite{wu2022state}.

The necessity of interpretability in multimodal recommendation is evident from three perspectives. From the user perspective, it reveals the modality driving a recommendation and lets users verify alignment with their preferences: natural-language explanations from user and item embeddings help patients understand treatment recommendations \cite{liang2025enhancing}, explainable course recommendation boosts learner engagement \cite{cen2025egrec}, and feature-level attribution improves comprehension of recommendation outcomes \cite{abu2024experimental}. From the system perspective, it supports debugging and fault localization by exposing the decision path through feature attribution such as integrated gradients \cite{rychalska2020know}, enabling model diagnosis in short-video recommenders \cite{zhao2025multimodal} and visual explanations from knowledge-graph reasoning in graph-fusion frameworks \cite{zare2025graphfusion}. From the regulatory perspective, it is a prerequisite for compliant AI operation: differential-privacy recommendation quantifies feature influence via gradients to calibrate noise and satisfy privacy rules \cite{yang2025research}, and video recommenders must comply with negative-feedback explanation requirements \cite{chen2026top}.

\subsection{Application Domains and Taxonomy}

Multimodal explainable recommender systems have achieved application progress in domains including fashion recommendation \cite{wu2020visual,li2021attribute}, visual recommendation \cite{chen2019personalized}, video recommendation \cite{zare2025graphfusion,chen2026top}, e-commerce recommendation \cite{laenen2022learning}, and education recommendation \cite{cen2025egrec,tao2021multi}. This paper establishes a two-dimensional classification framework based on the timing and manner of interpretability generation: (1) Modality explanation: The recommendation model embeds interpretability mechanisms at the design stage, endowing the model structure itself with interpretability such that the recommendation decision process can be directly revealed without additional post-hoc processing; (2) Post-hoc explanation: After the recommendation model completes its predictions, external methods are employed to explain the already-generated recommendation results.

\subsection{Literature Search Strategy}

Studies at the intersection of multimodal recommendation and explainable recommendation have grown continuously in recent years. To systematically cover this body of work, this survey adopts the following literature search strategy.

\textit{Search terms}: Since this survey centers on two core concepts, ``multimodal recommendation'' and ``interpretability,'' the primary search terms combine ``multimodal recommendation'' with ``explainable'' or ``interpretable.'' To broaden coverage, related terms such as ``multimodal fusion'' and ``cross-modal representation'' are further paired with ``explainable,'' while ``counterfactual explanation'' and ``disentangled representation recommendation'' are added as supplementary terms, corresponding to the specific methods reviewed in this survey.

\textit{Search sources}: This paper first retrieves research papers containing the above search terms through Google Scholar, and subsequently expands the search scope of relevant literature through academic databases including IEEE Xplore, ACM Digital Library, SpringerLink, and ScienceDirect. 

\textit{Publication types}: This paper only includes peer-reviewed high-level publications related to multimodal explainable recommendation from international conferences and journals, with works of domain influence on the arXiv preprint platform serving as supplements.

Between 2018 and 2026, the number of research papers in the field of multimodal explainable recommendation shows a growing trend. The primary reasons lie in that multimodal fusion algorithms have demonstrated favorable effects in alleviating data sparsity and cold-start problems, and the demand for interpretability has received widespread attention from researchers in the practical application of recommender systems.

\subsection{Comparison with Existing Surveys}

In recent years, multiple important surveys have emerged in the fields of multimodal recommendation and interpretability. These surveys have reviewed relevant research progress from different perspectives, providing an important foundation for this paper. However, none of the existing surveys systematically focuses on the interpretability issues of multimodal recommendation, exhibiting the characteristic of ``single-dimensional coverage with gaps in cross-domain areas.''

From the perspective of research themes, existing surveys fall into two lines. Surveys on multimodal recommender systems focus on fusion techniques and performance optimization, organizing existing work around feature extraction, encoders, modality fusion, and loss functions \cite{lopez2025survey,xu2026survey}; however, interpretability is only briefly mentioned as a future direction, without analyzing multimodal-specific dimensions such as modality alignment and modality contribution \cite{liu2024multimodal}. By contrast, surveys on unimodal explainable recommendation provide in-depth analyses of attention mechanisms and feature attribution and trace the evolution of interpretability methods \cite{sun2024review}, yet they assume unimodal input data and thus overlook the interpretability challenges arising from cross-modal interaction.

From the perspective of technical frameworks, three limitations remain. First, classification schemes are typically built on architectural components such as encoders and fusion layers \cite{zhou2026comprehensive,xu2026survey}, without delineating interpretability types. Second, multimodal-specific dimensions, including modality alignment, modality contribution, intra-modal, and inter-modal explanation, are largely absent, even though they distinguish multimodal explainable recommendation from unimodal scenarios. Third, coverage of emerging methods, such as knowledge-graph reasoning paths and LLM-generated explanations, remains limited \cite{lopez2025survey}. Moreover, existing surveys are relatively weak in terms of
evaluation frameworks, lacking systematic reviews of metrics, datasets, and benchmark experiments.

\subsection{Contributions of This Survey}

This paper presents the first systematic survey of interpretability techniques for multimodal recommendation, filling the research gap in this interdisciplinary field. Compared with existing surveys, the contributions of this paper are reflected in four aspects: (1) Based on the timing and manner of interpretability generation, this survey establishes a two-dimensional classification system that divides existing methods into modality explanation and post-hoc explanation, and constructs a four-level progressive hierarchy for each; (2) From objective quantification to subjective evaluation, this survey synthesizes evaluation metrics spanning dimensions such as modality contribution and counterfactual explanation quality, and classifies and reviews datasets according to application domain and data source; (3) Unique technical challenges faced by interpretability in multimodal scenarios are analyzed, including the modality alignment black box, the difficulty of fusion weights adapting to personalized preferences, and reliance on weakly supervised annotations, and response paths from cutting-edge progress such as information-theoretic decomposition and LLM evidence grounding are reviewed; (4) Unsolved problems are summarized, including the lack of unified standards for verifying the faithfulness of generated explanations and bottlenecks in counterfactual search efficiency, and future directions are envisioned, such as explanation as evidence and multimodal latent preference discovery, etc.

\subsection{Organization of the Survey}

This paper provides a comprehensive survey of existing research on multimodal explainable recommender systems. First, the basic concepts, research significance, and application scenarios of multimodal explainable recommender systems are introduced. Second, a classification framework based on the timing and manner of interpretability generation is established, dividing existing methods into two major categories: modality explanation and post-hoc explanation. In the modality explanation section, a systematic review is conducted from perspectives including modality alignment explanation, modality contribution explanation, intra-modal explanation, and inter-modal explanation; in the post-hoc explanation section, methods including feature importance analysis, multimodal LLM-generated explanations, knowledge graph reasoning paths, and counterfactual explanation are covered. Furthermore, evaluation methods, commonly used datasets, and evaluation metrics for multimodal explainable recommendation are summarized. Finally, the shortcomings of existing research are analyzed, and future research directions are pointed out, such as explanation as evidence and multimodal latent preference discovery, etc.

\section{Preliminaries}

\subsection{Definitions and Taxonomy of Explainability}

Interpretability denotes the ability of a model to clarify its decision-making process and outcomes for human users; in recommender systems, it is specifically the ability to answer ``why this item is recommended'' \cite{guo2023towards}. This notion involves three dimensions. First, decision transparency requires the model to expose the reasoning path from input features to recommendation results, and multimodal systems must further reveal how text, images, and audio jointly shape the decision \cite{alamdari2025edgex}. Second, result understandability concerns whether explanations can be grasped and accepted by users, and interpretability has been shown to raise user trust in recommender systems \cite{liang2025enhancing}. Third, process traceability allows users to follow how a recommendation decision is formed, which also improves learner engagement in educational recommendation \cite{cen2025egrec}.

\subsection{Modality Types and Features}

Each type of modality processed by multimodal recommender systems possesses a hierarchical semantic structure, primarily including four categories: Visual modality features have a hierarchical structure: low-level features (pixels, color) $\rightarrow$ mid-level features (edges, regions) $\rightarrow$ high-level features (objects, style) \cite{chen2018visually}. Visual attention mechanisms can identify salient regions to generate visual explanations \cite{chen2019personalized,chen2018visually}; the textual modality ranges from surface-level lexical features to deep semantic vector representations, including keyword features, topic features, sentiment features, and semantic vectors \cite{ye2025harnessing,geng2023vip5}, and textual features are often jointly modeled with visual features to achieve image-text consistency alignment \cite{wu2020visual,geng2022improving}; the audio modality progresses from low-level acoustic features to high-level semantic features, including spectrum, MFCC, rhythm, and so on \cite{zare2025graphfusion,yang2025research}; the video modality is a composite modality of image sequences and audio, requiring simultaneous processing of spatial and temporal information \cite{chen2026top}.

Multimodal fusion is divided into three types according to the timing of fusion: feature-level fusion (early fusion) concatenates or performs weighted fusion of different modality features at the feature level \cite{sagar2020pai}; model-level fusion (intermediate fusion) fuses modality features at the model level and can display the interaction process of each modality \cite{zare2025graphfusion}; decision-level fusion (late fusion) allows each modality to make predictions independently, clearly demonstrating the decision weights of each modality \cite{sang2020context}.

\subsection{Types of Recommendation Tasks}

Classified by explanation demand, different recommendation scenarios have different requirements for interpretability. In high-risk decision recommendation scenarios, users require detailed and accurate recommendation justifications, and the demand for interpretability is the highest \cite{liang2025enhancing}; in consumer decision recommendation scenarios, users need to understand the degree of match between product characteristics and their own needs \cite{laenen2022learning}; in entertainment content recommendation scenarios, interpretability can enhance user trust in recommender systems. According to the presentation form of explanations, recommendation explanations are divided into three categories: visual explanations, textual explanations, and structured explanations. Visual explanations present the basis for recommendations by highlighting image regions \cite{chen2019personalized}; textual explanations present recommendation reasons through natural language, keyword highlighting, and other means \cite{laenen2022learning}; structured explanations present reasoning chains through knowledge graph paths \cite{cen2025egrec}.

\section{Modality-Level Explainability}


\input{table_modality_explainability}

\subsection{Modality Alignment Explanations}

Modality alignment explanation demonstrates how textual, visual, audio, and other modal features are mapped into a unified semantic space, revealing the process by which cross-modal associations are established \cite{alamdari2025edgex,yang2025explainable}. As the foundation of multimodal recommendation, its interpretability directly influences users' understanding of recommendation decisions.

Cross-modal feature alignment attributes recommendation results to explicit alignment operations, typically implemented through contrastive loss or weakly supervised alignment. EARec aligns text, image, and behavior modalities in parallel to shared interpretable anchors, and fine-tunes multimodal large models with generative tasks so that the alignment process remains explainable and traceable \cite{yang2025explainable}. EdgeX-MMFRec further introduces a cross-modal alignment (CMA) layer that maps pretrained text, image, and audio encoders into a unified semantic space, together with an explanation-fidelity objective that keeps the generated explanations consistent with the alignment process \cite{alamdari2025edgex}.

Alignment loss design first optimizes the generation of shared interpretable anchors through a generative alignment task, guiding the model toward meaningful cross-modal relationships \cite{yang2025explainable}. Contrastive loss complements this by pulling similar modal features together and penalizing image-text inconsistency, which makes the learned alignment more interpretable \cite{geng2022improving}. Under weakly supervised conditions, the MDR model attaches an independent attribute classifier to each disentangled latent dimension, anchoring dimensions to semantics such as color and shape through cross-entropy loss and limited annotations, while imposing orthogonal constraints to keep attribute factors mutually independent; this aligns disentangled representations with semantic attributes at low annotation cost without sacrificing alignment quality \cite{wang2021multimodal,cheng2019mmalfm}.

Disentangled representation alignment decomposes multimodal representations into visual-dominant attributes, text-dominant attributes, and shared attributes, clearly demonstrating the independent contributions of different modalities to recommendations \cite{laenen2022learning}. Multimodal Disentangled Representation learning (MDR) decomposes images into independent generative factors (e.g., color, neckline shape) through $\beta$-VAE, such that each dimension corresponds to an interpretable semantic attribute \cite{wang2021multimodal}. The DGVAE model utilizes CLIP to convert images into textual descriptive words, achieving text-bridged modality alignment, and aligns interaction-derived and text-derived disentangled representations through mutual information maximization, enabling user preferences to be intuitively displayed via word clouds \cite{zhou2024disentangled}.

Alignment visualization presents cross-modal associations through approaches such as keyword-region correspondence displays and disentangled factor alignment demonstrations \cite{laenen2022learning}. The VTJEI model employs a bidirectional two-layer adaptive attention review model to highlight keywords as textual explanations, while simultaneously highlighting image regions as visual explanations through a review-driven visual attention model \cite{wu2020visual}. Research shows that generated image visualizations can intuitively demonstrate alignment effects and enhance user understanding and trust \cite{geng2022improving}.

\begin{figure*}[t]
\centering
\includegraphics[width=6.0in]{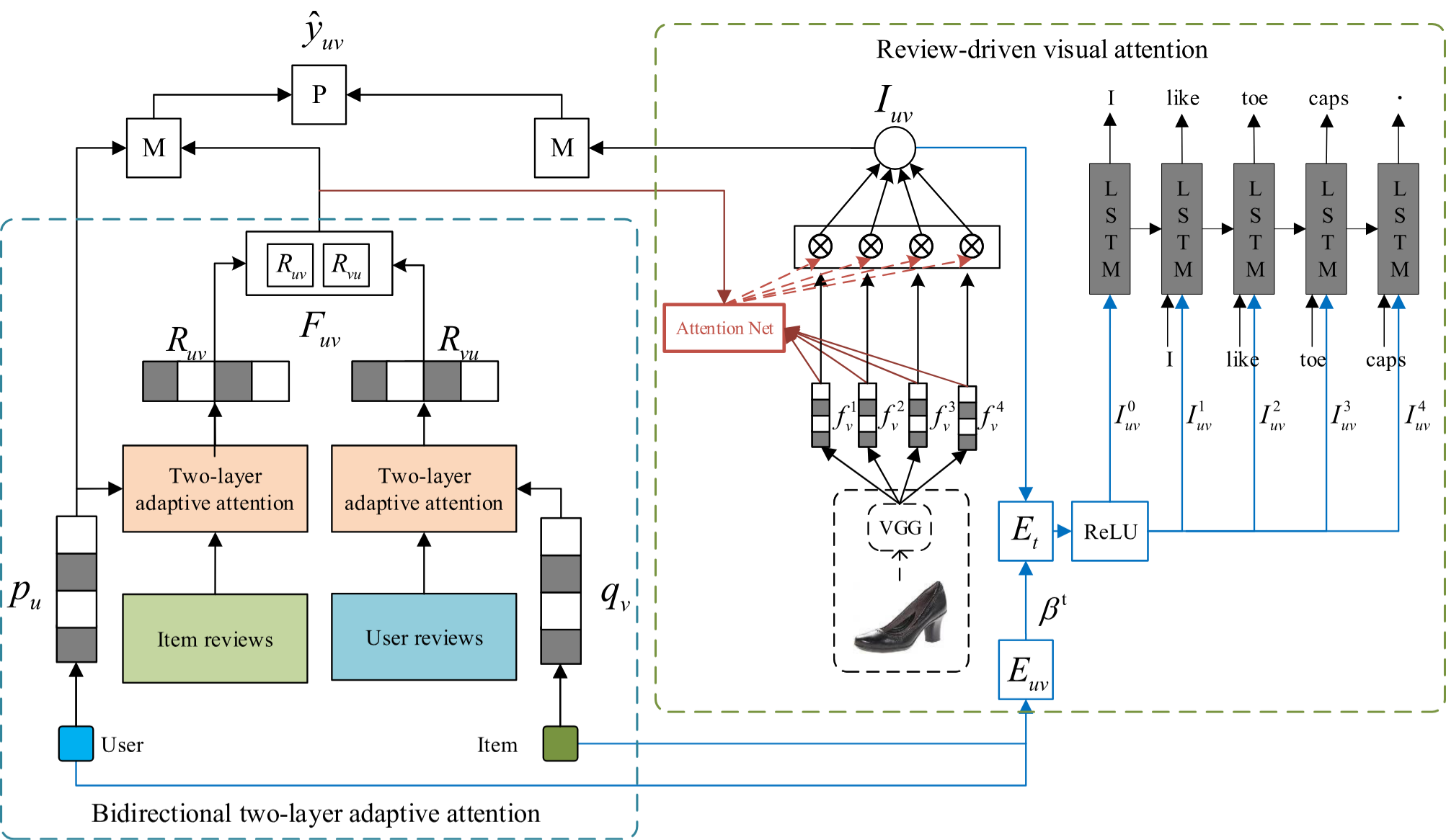}
\caption{The overall framework of the VTJEI model. The dashed box on the left is a bidirectional two-layer adaptive attention model, which can obtain the user's preference for the target item; the dashed box on the right is a review-driven visual attention model, which can obtain a personalized image representation of the item \cite{wu2020visual}.}
\label{fig:vtjei}
\end{figure*}

\subsection{Modality Contribution Explanations}

Modality contribution explanation demonstrates the proportional contribution of each modal feature to recommendation results, helping users understand the multimodal information fusion process \cite{zhao2025multimodal,zare2025graphfusion}. Contribution explanation constitutes the core content of multimodal explainable recommendation, directly answering the question of ``which modality influenced the recommendation.''

Modality weight allocation computes importance scores for the text, image, audio, and other modalities through attention mechanisms. AM-Bi-LSTM adopts an adaptive multimodal bidirectional LSTM with a modality attention module that separately scores the text, image, and identifier modalities, and uses the share of each score in the final recommendation as the explanation basis \cite{ohtomo2024bi}. GraphFusion-HRL instead builds a co-attention graph fusion module that aligns textual, visual, and structural features with a dual-stream attention mechanism and outputs the weight distribution of every modality \cite{zare2025graphfusion}. A mixture-of-experts network takes a different route: its gating mechanism dynamically assigns fusion weights to the structural, textual, and visual modalities, and the gating weights directly quantify how much each modality and attribute contributes to the recommendation, offering finer-grained dynamic quantification of modality contribution \cite{liu2025multimodal}.

The concrete implementation of this weight allocation is the modality confidence estimation module, which computes a reliability score for each modality with a single-layer attention network. The resulting weights directly measure how much each modality contributes to the recommendation decision, thereby giving modality contribution explanation a quantifiable computational basis. The modality confidence weight is defined as follows:
\begin{equation}
w_i^{(m)}=\sigma\!\left(\mathbf{u}^{\top}\tanh\!\left(\mathbf{V}^{(m)}\hat{\mathbf{e}}_i^{(m)}\right)\right),\qquad \sum_m w_i^{(m)}=1,
\label{eq:modality-confidence}
\end{equation}
where $w_i^{(m)}$ is the confidence weight of item $i$ on the $m$-th modality, with $m\in\{s,t,v\}$ indexing the structural, textual, and visual modalities; $\sigma(\cdot)$ is the activation function; $\mathbf{u}$ is the user preference vector; $\mathbf{V}^{(m)}$ is the learnable transformation matrix for the $m$-th modality; and $\hat{\mathbf{e}}_i^{(m)}$ is the $\ell_2$-normalized embedding of item $i$ on the $m$-th modality, obtained by projecting the raw modality feature into the shared latent space. The weights are normalized to satisfy $\sum_m w_i^{(m)}=1$ \cite{zare2025graphfusion}.

Attention weight visualization displays cross-modal correlations through co-attention matrices, where modal weights directly reflect the importance of each modality \cite{zare2025graphfusion}. Research demonstrates that visualization of learned hash codes and explicit semantic-attribute scores provides interpretability for outfit compatibility \cite{guan2022multimodal}. A multimodal foundation model-driven interest modeling framework visualizes the degree of contribution of each modality to recommendations through attention weights, revealing the basis for model decisions \cite{zhao2025multimodal}.

Contribution quantification calculates feature importance weights through gradients, decomposing the total reward into dimensions such as engagement, diversity, and modal confidence \cite{cen2025egrec}. GraphFusion-HRL decomposes the total reward into three components---engagement, diversity, and modal confidence---with dynamic weights illustrating how the model balances multiple objectives \cite{zare2025graphfusion}. Research on differential privacy protection mechanisms computes the influence of features on recommendation results through gradients and dynamically adjusts noise, balancing privacy protection and recommendation effectiveness\cite{yang2025research}. A theory-based multimodal deep learning architecture uses a stick-breaking mechanism to measure the asymmetrical influence of image and text descriptions on each recommendation decision, clearly distinguishing whether a purchase is driven by visual information or textual information. The study decouples the influence of visual and textual features on user preferences by injecting user reviews, providing a theoretical quantification method for modality contribution explanation \cite{liu2026unveiling}.

\subsection{Intra-Modality Explanations}

Intra-modal explanation reveals how features within a single modality influence recommendation decisions, serving as a critical link for understanding the microscopic mechanisms of multimodal recommender systems \cite{wu2020visual}. Unlike modality contribution explanation, which asks ``which modality matters more,'' intra-modal explanation delves inside one modality to identify the specific features, attributes, or regions that decisively shape the decision. Across the visual, textual, and audio modalities, researchers have developed distinctive methods that together form a multi-level explanatory system ranging from coarse to fine and from superficial to profound.

Visual feature explanation locates key regions in images through visual attention, providing intuitive visual evidence for recommendation decisions. The influence of visual features can be quantified at different granularities: AM-Bi-LSTM computes a visual importance degree via within-modality matching \cite{ohtomo2024bi}, while BiHGH explains item compatibility through explicit attribute-level semantic scores \cite{guan2022multimodal}. At the region level, the VECF and Re-VECF models learn visual attention supervised by both implicit feedback and textual reviews, so that different users attend to different regions of the same product image, revealing fine-grained personalized visual preferences \cite{chen2018visually}. Beyond localization, visual explanation must establish semantic correspondence with user preferences. Review-driven visual attention methods extract preference semantics from users' historical reviews to guide attention learning, ensuring that highlighted regions correspond to preference words such as color, neckline, and cuff \cite{chen2019personalized}.

A user-adaptive filtering module further personalizes the process by dynamically generating filters whose activation maps expose the poster regions the model attends to, for example dark and tense regions for horror-film users versus bright and vibrant regions for Disney users \cite{torres2023multimodal}. Ante-RNN aligns image regions with textual words and applies dynamic context attention to recent browsing behaviors, allowing explanations to adapt as user interests drift \cite{liu2020dynamic}. Visual attribute disentanglement pushes the granularity further by decomposing visual features into interpretable semantic dimensions such as color, material, and style, each carrying explicit meaning \cite{ohtomo2024bi}. Weakly supervised attribute localization learns attribute-specific regions with a predefined attribute set, verifying representation-attribute correspondence at reduced annotation cost \cite{li2021attribute}.

Textual feature explanation encodes text into interpretable semantic factors, extracting preference features from user reviews and product descriptions. Large-model-based methods transform each item's visual and textual data into a unified textual description, so that user preferences are extracted from one consistent source, presenting the textual basis for recommendations \cite{ye2025harnessing}. Studies of user reviews distinguish visible preferences, which users perceive directly from images such as clothing color, style, and sleeve type, from invisible preferences, such as material, quality, comfort, and craftsmanship, which are accessible only through textual reviews. Large model intent prompt learning goes finer by clustering semantic intent prompts such as price, color, and scent from user reviews, where each prompt corresponds to one interpretable user intent dimension \cite{jian2026large}.

Semantic factor analysis offers a complementary route: it encodes items into interpretable semantic factors spanning shape, color, pattern, and style dimensions, revealing the meaning of each \cite{li2021attribute}. By recursively summarizing a user's historical interactions into textual preference descriptions, the model renders its implicit interest states readable in text form, making preferences traceable \cite{ye2025harnessing}. Textual explanation also serves beyond the textual modality: the Re-VECF model uses preference words in review text to constrain visual attention learning, ensuring that highlighted image regions align with visual features mentioned in user reviews \cite{chen2018visually}.

Audio feature explanation works out the attention weight of the audio modality through modality-level attention, bringing to light the audio modality's contribution to recommendation results \cite{zhao2025multimodal}. In short video recommendation, audio features include rhythm, speech rate, and speaker characteristics, and their fused weights vary across users and content. Through attention weight visualization, research can show the dynamic weight of the audio modality when visual, textual, and audio modalities fuse together, revealing variations in modality dominance across user scenarios or content types.

\subsection{Inter-Modality Explanations}

Inter-modal explanation reveals the interactive relationships among different modalities, demonstrating the process by which multimodal collaboration influences recommendations \cite{sagar2020pai,sang2020context}. Inter-modal explanation answers the question of ``how different modalities cooperate,'' constituting an essential component of multimodal explainable recommendation.

Cross-modal complementarity investigates the compensatory mechanisms of other modalities when one modality is missing. Multimodal recommendation systems leverage cross-modal cues for richer collaborative reasoning. The meta-path random walk method can construct homogeneous video graphs and learn shared multimodal video features, enabling multimodal collaborative reasoning \cite{sang2020context}.

Attribute-level modality interaction explanation analyzes modal interactions at the fine-grained attribute level, locating the specific attributes that lead to disharmony \cite{sagar2020pai,li2021attribute}. PAI-BPR, for example, adopts a multi-task attribute classification network to extract five readable attribute categories---texture, style, fabric, silhouette, and component details---for personalized clothing recommendation, achieving attribute-level interpretability \cite{sagar2020pai}. In the audio domain, an ontology-guided music recommendation framework organizes singers, genres, and emotions into a structured ontology and re-ranks items by ontology compatibility scores, thereby providing concept-level explanations such as ``similar in sound and sharing a genre'' for high-level semantic associations across modalities \cite{rumiantcev2026ontology}. A personalized display method further pairs selected personalized image sets via CLIP alignment with matching natural language explanations, overcoming the information poverty of purely textual explanations \cite{yan2023personalized}.

As described above, PAI-BPR extracts multi-dimensional readable attributes for clothing images through a multi-task attribute classification network, and the construction process of this attribute representation concatenation serves as the computational foundation for inter-modal attribute-level interaction. From the perspective of interpretability, the concatenation of semantic attribute vectors not only achieves fine-grained feature organization within the visual modality, but also constitutes the explanatory basis for subsequent cross-attribute compatibility reasoning. The semantic attribute representation concatenation is defined as follows:
\begin{equation}
\mathbf{f}_i=[\mathbf{f}_i^{1};\mathbf{f}_i^{2};\ldots;\mathbf{f}_i^{Q-1};\mathbf{f}_i^{c}],
\label{eq:attribute-concat}
\end{equation}
where $\mathbf{f}_i$ is the semantic attribute representation of the $i$-th fashion item, and $[\,\cdot\,;\cdots\,;\cdot\,]$ denotes vector concatenation; $\mathbf{f}_i^q=h^q(I\,|\,\theta^q)$ is the feature vector of the $q$-th attribute extracted from the item image $I$ by the attribute classification network $h^q$, with $\theta^q$ denoting the parameters of $h^q$ and $q\in\{1,2,\ldots,Q-1\}$ indexing the visual attribute types (texture, style, fabric, shape, and part details); and $\mathbf{f}_i^c$ is the one-hot semantic attribute representation derived from the category label of the item \cite{sagar2020pai}.

Graph reasoning path explanation leverages knowledge graphs to provide structured explanations, generating recommendation path explanations through meta-path guided reasoning \cite{cen2025egrec}. The EGRec model constructs a course knowledge graph and provides recommendation rationale through relational paths in the knowledge graph (e.g., ``Course B $\rightarrow$ prerequisite $\rightarrow$ Course A''), explaining the logical connections between courses \cite{cen2025egrec}. The MKRLN model models recommendation as a reinforcement learning process, generating reasoning paths on a multimodal knowledge graph and achieving dual explanations of knowledge paths and image paths \cite{tao2021multi}.

Fusion strategy explanation analyzes inter-modal interactions from the perspective of fusion timing \cite{sagar2020pai,sang2020context}. Early fusion operates at the feature level, explaining the fused representation; intermediate fusion operates at the model level, explaining the fusion process; late fusion operates at the decision level, explaining the decision weights of each modality. The MRG model designs a fusion gate as a soft attention mechanism that dynamically weighs the contributions of affective features and visual features at each time step of the LSTM, belonging to an intermediate fusion strategy \cite{truong2019multimodal}. The MMCT model softly balances affective and visual features through time-varying gating, where the magnitude of the $\beta$ value directly reflects the model's emphasis on affective versus visual features \cite{liu2023multimodal}. Research suggests that different fusion strategies exhibit distinct explanatory characteristics, requiring the selection of appropriate strategies according to the application scenario.

\subsection{Relationships and Distinctions}

The four levels of modality explanation present a progressive relationship in analytical granularity, from macroscopic to microscopic and from single-modality to interactive. Modality alignment focuses on mapping different modal features into a unified semantic space \cite{alamdari2025edgex,yang2025explainable}. Modality contribution explanation quantifies the importance ranking of each modality through attention weights or gating scores \cite{zhao2025multimodal,ohtomo2024bi}. Intra-modal explanation dissects which specific features within a single modality play a decisive role in recommendation decisions \cite{wu2020visual,chen2018visually}. Inter-modal explanation concerns how different modalities cooperate, complement each other, and produce synergistic effects \cite{sagar2020pai,sang2020context}. These levels are interdependent rather than parallel: alignment is a prerequisite for contribution explanation, contribution points the direction for intra-modal analysis, and the fine-grained attributes revealed intra-modally in turn form the basis for inter-modal interaction analysis \cite{sagar2020pai,li2021attribute}, together constituting a complete chain from spatial alignment through weight allocation and feature dissection to interaction modeling.

Each level also has its limitations, which recent work has begun to address. The black-box nature of the alignment process limits verifiability \cite{yang2025explainable}; to counter this, an information synergy module decomposes multimodal information from an information-theoretic perspective into four categories---image-unique, text-unique, redundant, and synergistic---endowing synergistic information with an explicit definition and quantification method for the first time \cite{li2026prism}, while ExplainRec maps user--item interactions and preference attributions into the LLM semantic space through instruction tuning \cite{ma2025explainrec}, and a dynamic adapter method achieves online multimodal alignment to accommodate preference changes \cite{ma2025bridging}. Fixed fusion weights struggle to adapt to personalized preferences \cite{ohtomo2024bi,li2026prism}; a mixture-of-experts network addresses this by dynamically allocating modal fusion weights through a gating mechanism \cite{liu2025multimodal}, and channel-level interaction scores quantify the independent contributions of textual and visual channels respectively \cite{jian2026large}. Reliance on weakly supervised annotation constrains the scalability of intra-modal explanation \cite{wu2020visual,wang2021multimodal}; large model intent prompt learning mitigates this by automatically clustering semantic intent prompts such as price and color from user reviews, reducing dependence on human annotation \cite{jian2026large}. Finally, the modeling of synergistic information lacks a unified framework \cite{sagar2020pai,li2026prism}; ontology-guided music recommendation addresses this by establishing high-level semantic associations between modalities through structured ontologies \cite{rumiantcev2026ontology}.

\section{Post-Hoc Explanations}

\input{table_posthoc_explainability}

\subsection{Feature Importance Analysis}

Feature importance analysis generates explanations by quantifying the contribution of each feature to the prediction result \cite{rychalska2020know}. As the most classical post-hoc explanation approach, it is characterized by strong generalizability and ease of comprehension \cite{lee2026integrating}.

Gradient analysis methods employ integrated gradients to compute attribution values, attributing recommendation results to specific input features. Studies have demonstrated that integrated gradients can resolve the saturation problem inherent in traditional gradient methods and deliver more stable attribution outcomes. The interpretability method for EMDE leverages integrated gradients to calculate attributions over the input sketch for the output buckets of the highest-scoring item, and through a decoding procedure that averages bucket-level attributions, it obtains an approximate attribution score for each input item and thus attributions for individual items across modalities \cite{rychalska2020know}.

Feature importance ranking identifies the most important attributes through a feature-importance architecture, helping users intuitively grasp which attributes drive each prediction \cite{sagar2020pai}. Research indicates that decision-tree visualization, with attributes as tree nodes, can present feature importance in an easily readable form. A cross-domain medical recommendation model uses GPT-2 to generate natural language explanations, taking user and item embedding vectors as continuous prompt inputs to produce personalized explanations \cite{liang2025enhancing}.

The primary strength of feature importance analysis lies in its generalizability, rendering it applicable to virtually all types of recommendation models. Nevertheless, post-hoc explanation may suffer from faithfulness issues, meaning that the explanation may not align with the actual decision-making process of the model. Research suggests coupling the explainable agent with the recommender rather than running a detached post-hoc generator, so that the fidelity loss enforces faithful explanations and improves both accuracy and trust \cite{alamdari2025edgex}.

\subsection{Multimodal LLM-Generated Explanations}

Multimodal LLM-generated explanation leverages large language models to produce natural language explanations, offering a new technical pathway for recommendation explanation generation \cite{ma2025explainrec,ma2025bridging}. LLM-driven explanations boast the advantages of natural expression and strong readability, and have garnered widespread attention in recent years.

LLM-driven natural language explanation generates textual descriptions of user preferences, providing natural language justifications for recommendation reasons. The MLLM-MSR model fuses image and text information into a unified textual description via a multimodal LLM and, through recursive user preference inference, produces preference descriptions that serve as recommendation rationales (e.g., ``the user enjoys Chinese cooking and humorous-style content'') \cite{ye2025harnessing}. The MMP-Refer framework feeds node profiles of retrieval paths as hard prompts and interaction subgraph encodings as soft prompts into an LLM; a Mixture of Experts (MoE) collaborative adapter maps structured information into LLM-comprehensible prompts, yielding fluent and targeted rationales such as ``this item is visually similar in style to items you have previously purchased'' \cite{pan2026mmp}. A medical cross-domain recommendation model generates explanations with GPT-2, aligning prompts and model parameters through a two-stage tuning strategy to produce patient-friendly output \cite{liang2025enhancing}. Research further shows that LLM-generated captions can describe the rationale behind each recommendation and help users understand why two items are associated \cite{zhang2026reasonrec}, enhancing the trust and acceptance of such explanations \cite{arnold2024explainable}.

\begin{figure*}[t]
\centering
\includegraphics[width=6.0in]{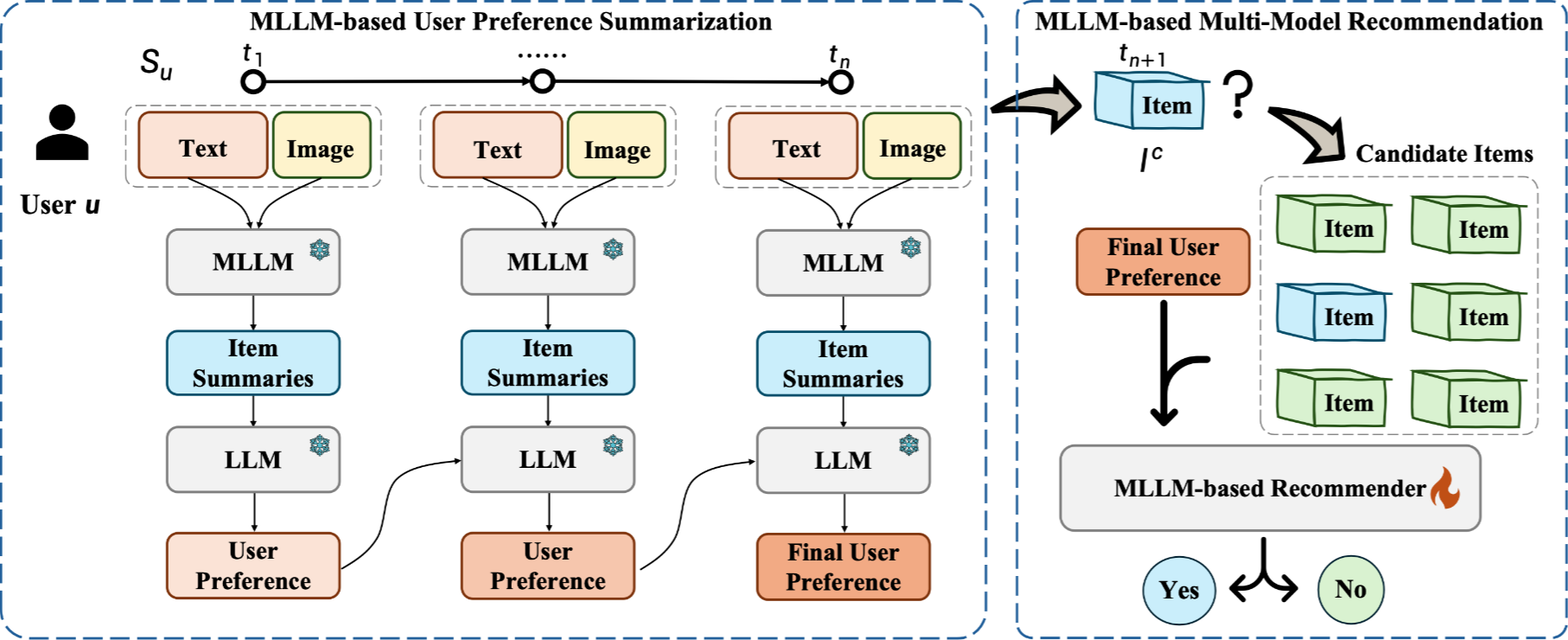}
\caption{The schematic framework of MLLM-MSR \cite{ye2025harnessing}.}
\label{fig:mllm-msr}
\end{figure*}

A second line of work couples explanation generation with the recommendation process itself. The LLM4Rec model employs an LLM as the backbone to jointly produce recommendations and explanations, deriving preference-based, similarity-based, and context-based rationales from latent factor analysis, embedding similarity, and contextual attention, respectively \cite{ma2025llm4rec}. ExplainRec feeds historical preference reasons to an LLM through preference attribution instructions so that a single model outputs both the prediction and its explanation; joint training over prediction, explanation, and consistency losses keeps the two intrinsically unified \cite{ma2025explainrec}. An evidence grounding approach extracts collaborative neighbors and textual and visual attributes as explicit evidence tokens, prompting a frozen LLM to generate evidence-based explanations whose consistency with the prediction basis is enforced by an evidence-removal faithfulness loss \cite{ma2025bridging}. Beyond generation, a reinforcement learning framework folds multimodal explanation quality into closed-loop optimization, rewarding recommendation relevance, explanation attractiveness, and multimodal consistency \cite{lin2026reinforcement}, while a reasoning-enhanced agent casts recommendation as chained observation, reasoning, and action steps, exposing intermediate reasoning with quantified uncertainty so that explanations coincide with the model's own reasoning process \cite{zhang2026reasonrec}.

Video content analysis-based explanation targets recommendation failure scenarios, generating personalized explanations for negative feedback reasons through multi-granularity video content analysis. The Agentic Explainable Negative Feedback framework, through multi-agent collaboration, extracts behavioral clues from historical data, performs multimodal video analysis, synthesizes information to predict user engagement, and generates explanations \cite{chen2026top}.

Multimodal information verbalization converts visual and textual information into unified item descriptions and textual preference summaries \cite{ye2025harnessing}. Research indicates that showcasing the fusion process of textual and visual information through attention mechanisms can enhance the credibility of LLM-generated explanations. The TourMLLM model, when generating tourism recommendation explanations, explicitly utilizes externally retrieved knowledge (e.g., images, knowledge graph triples, review summaries), ensuring that each tourism recommendation explanation is grounded in task-specific retrieved knowledge \cite{yamanishi2025tourmllm}. The MSRBench benchmark framework systematically evaluates integration strategies of large vision-language models in sequential recommendation, finding that models perform optimally when acting as rerankers and can simultaneously generate recommendation reasons \cite{zhou2025large}.

\subsection{Knowledge Graph-Based Reasoning Paths}

Knowledge graph reasoning path utilizes the structured information of knowledge graphs to generate interpretable recommendation paths \cite{zhang2025hkgat,tao2021multi}. Knowledge graphs provide rich semantic relationships that can support the generation of causal reasoning chains, representing an important technical pathway for post-hoc explanation.

Knowledge graph path visualization extracts user--item paths from the knowledge graph and renders them as knowledge subgraph visualizations with tools such as Neo4j \cite{zhang2025hkgat,cen2025egrec}. MMP-Refer performs heuristic search within the user--item interaction subgraph, building edge weights from hop count, node degree, and multimodal similarity to retrieve explanatory paths. In the resulting visualization, path connectivity links historically interacted items to the recommended item, node profiles present product names and attributes, and edge weights reflect the degree of multimodal similarity, so that users can intuitively follow the reasoning of ``why this item is recommended'' \cite{pan2026mmp}. HKGAT likewise extracts user--item paths, computes path interpretability scores, and generates Neo4j subgraph visualizations that display recommendation paths and their relationships \cite{zhang2025hkgat}. EGRec, in turn, justifies recommendations through relational paths such as ``Course B $\rightarrow$ prerequisite $\rightarrow$ Course A,'' making the logical connections between courses explicit \cite{cen2025egrec}.

Entity-relationship explanation displays association paths among users, items, and attributes, generating semantic path explanations \cite{zhang2025hkgat,cen2025egrec}. Research indicates that typed nodes (courses, chapters, knowledge points) and typed edges (prerequisite, similarity, containment) enrich the semantics of path explanations, while causal evidence can be supplied from both knowledge and images \cite{tao2021multi}. The PAI-BPR model analyzes modal interactions at the attribute level, identifying specific attributes that cause disharmony through attribute-pair correlation analysis and providing fine-grained explanations \cite{sagar2020pai}.

Semantic reasoning chain refers to the inference chain from knowledge to recommendation, demonstrating the complete semantic reasoning process of recommendation decisions \cite{zhang2025hkgat,cen2025egrec}. The MKRLN model models recommendation as a reinforcement learning process, generating reasoning paths on a multimodal knowledge graph, where each hop uses an attention mechanism to filter neighbor nodes in which the user is truly interested, forming a hierarchical attention path \cite{tao2021multi}. The first-hop attention filtering weight is shown in Eq.~\eqref{eq:mkrln-attention}:
\begin{equation}
\chi_1=\mathrm{softmax}(e_t,e'_t)=\frac{\exp(e_t,e'_t)}{\sum_{e''_t\in N_{e_t}}\exp(e_t,e''_t)},
\label{eq:mkrln-attention}
\end{equation}
where $\chi_1$ is the normalized first-hop attention weight, $\mathbf{e}$ is the embedding of the current entity, $\mathbf{e}'$ is the embedding of a neighbor entity, $\mathcal{N}(\mathbf{e})$ is the set of neighbors of $\mathbf{e}$ in the knowledge graph, and $\exp(\mathbf{e},\mathbf{e}')$ is the inner-product score of the two entities. The softmax function normalizes the scores over all neighbors $\mathbf{e}''\in\mathcal{N}(\mathbf{e})$ so that the weights sum to one \cite{tao2021multi}.

Knowledge graph-enhanced explanation improves interpretability by injecting knowledge into recommendation models, expanding item semantic descriptions, and optimizing ranking for interpretability. HKGAT, for instance, designs a path-mining-based explainability quantification module that computes user--item path scores and re-ranks the top-10 recommendation results by interpretability score, improving the reasonableness of explanations \cite{zhang2025hkgat}. Knowledge injection further broadens the semantic description of items, making recommendation results more interpretable \cite{tao2021multi}. The principal strength of knowledge graph reasoning paths lies in their structured and causal character; however, constructing knowledge graphs is costly, and application is limited in scenarios where structured knowledge is scarce. Augmenting path reasoning with explicit reasoning over knowledge and images mitigates this limitation, lowering the application threshold and broadening the applicability of knowledge graph explanation \cite{tao2021multi}.

\subsection{Counterfactual Explanations}

Counterfactual explanation explains recommendation decisions by constructing counterfactual scenarios that address ``how would the recommendation result change if a certain feature were different,'' revealing the intrinsic logic of recommendation outcomes from the perspectives of causal intervention and effect quantification \cite{liu2026rice,jain2024counterfactual}.

Counterfactual scenario construction creates counterfactual scenarios that flip recommendation results by modifying interaction histories, perturbing image embeddings, or selecting similar unrecommended items; the core challenge lies in identifying the minimal feature change that alters the recommendation decision. The RICE model uses concatenated textual and visual embeddings to compute multimodal similarity for guiding candidate generation, and under frozen validation, iteratively optimizes through beam search to find minimal interaction modifications \cite{liu2026rice}. The CAVIAR model searches for minimal perturbations in the image embedding space that cause recommendation flipping, and uses CLIP to map the perturbations to textual aspect classes \cite{jain2024counterfactual}. The MACE model filters counterfactual candidates based on CLIP visual similarity and low recommendation scores, and combines BLIP-2 visual descriptions to support multi-aspect comparison \cite{zhou2025multimodal}.

Causal intervention debiasing performs counterfactual intervention through a causal inference framework to achieve debiasing and effect quantification; its core lies in distinguishing genuine preference from bias factors in recommendation results. Originally developed for multimodal fact verification, the Harnessing Counterfactual Reasoning model simulates the removal of a certain modality of evidence to construct counterfactual scenarios, introducing average treatment effect (ATE) and individual treatment effect (ITE) to quantify modal causal contributions \cite{li2025harnessing}. The Explainable Multimodal Recommendation model constructs a structural causal model, blocking spurious causal paths through do-calculus and generating counterfactual masks to distinguish bias from genuine preference \cite{zhou2026explainable}. 

Attribution analysis provides the basis for ranking key features in counterfactual scenarios by quantifying the marginal contribution of each feature; its core lies in comparing actual inputs against counterfactual baselines. For the compressed sketch representation of EMDE, Rychalska et al. apply integrated gradients to the output buckets of the highest-scoring items and then decode the sketch-level attributions back to individual input items, enabling fine-grained multimodal attribution even under compressed representations \cite{rychalska2020know}. DeepSHAP quantifies the positive and negative effects of text, image, and rating features on the basis of Shapley values, its logic amounting to the counterfactual question ``how would the prediction change if this feature were removed'' \cite{lee2026integrating}. Average treatment effect and individual treatment effect extend attribution from the feature level to the modality level from a causal inference perspective \cite{li2025harnessing}.

Cross-modal semantic mapping translates visual counterfactual changes into human-readable semantic explanations; its core addresses the problem that visual perturbations are unintelligible to users. The CAVIAR model uses CLIP to map image embedding perturbations to textual aspect classes, making it understandable ``which visual aspect was modified to cause the recommendation to flip'' \cite{jain2024counterfactual}. The MACE model converts item images into descriptive text through BLIP-2, enhancing the LLM's understanding of visual differences to generate structured comparative explanations \cite{zhou2025multimodal}. The Explainable Multimodal Recommendation model decouples textual and visual features through cross-modal contrastive learning, providing independently manipulable semantic dimensions for counterfactual intervention \cite{zhou2026explainable}.

\subsection{Relationships and Distinctions}

Feature importance analysis, multimodal LLM-generated explanation, knowledge graph reasoning path, and counterfactual explanation all intervene in the explanation process after recommendation decisions are completed, yet they diverge in the depth of causal inference and in readability. Feature importance analysis attributes recommendation results to numerical contribution rankings through integrated gradients and Shapley values, theoretically solid but weak in readability \cite{rychalska2020know,lee2026integrating}. LLM-generated explanations output natural language rationales with high user acceptance, yet whether the generated content is consistent with the model's true decision-making remains unresolved \cite{ma2025explainrec,ma2025bridging}. Knowledge graph reasoning paths trace recommendation evidence along structured relational chains, offering clear logic, though constrained by graph coverage and path search efficiency \cite{zhang2025hkgat,sang2020context}. Counterfactual explanation constructs hypothetical scenarios through causal intervention and carries the strongest causal inference capability, but searching for valid counterfactual samples in high-dimensional multimodal spaces is computationally expensive, and visual perturbations often depend on external models to become readable semantics \cite{liu2026rice,jain2024counterfactual}.

These four methods are not isolated from one another; rather, they intersect and complement each other. Feature importance analysis provides a ranking basis for screening key features in counterfactual explanation \cite{rychalska2020know}, multimodal retrieval paths supply structured evidence chains for LLM-generated explanations \cite{pan2026mmp}, and counterfactual reasoning endows feature importance with causal validation capability \cite{li2025harnessing}. Combining multiple post-hoc explanation methods therefore often yields greater value than any single method alone.

Each method also faces distinct difficulties. The consistency between LLM-generated explanations and the model's true decisions is difficult to guarantee, and early methods risked ``fabricating out of thin air'' \cite{ma2025explainrec,ma2025bridging}; to relieve this trust crisis, evidence grounding extracts collaborative neighbors and textual and visual attributes as explicit evidence tokens to constrain the LLM toward evidence-based explanations \cite{ma2025bridging}, a reinforcement learning framework folds explanation quality into closed-loop optimization through rewards for recommendation relevance and multimodal consistency \cite{lin2026reinforcement}, and the reasoning-enhanced agent turns its intermediate reasoning steps into traceable explanatory evidence while quantifying its own predictive uncertainty \cite{zhang2026reasonrec}. Counterfactual explanation was long confined to pure collaborative filtering scenarios and unable to handle multimodal information, while visual perturbation results were unintelligible to users \cite{liu2026rice,jain2024counterfactual}; RICE therefore introduces textual and visual embedding similarity into counterfactual candidate generation \cite{liu2026rice}, CAVIAR maps image perturbations to textual aspect classes with CLIP so that users understand ``which visual aspect was modified to cause the recommendation to flip'' \cite{jain2024counterfactual}, and MACE leverages BLIP-2 to upgrade counterfactual explanation from single-point interpretation to multi-candidate comparative analysis \cite{zhou2025multimodal}. Knowledge graph paths, finally, remain constrained by graph coverage \cite{zhang2025hkgat,sang2020context}; temporal knowledge graph methods address this by bringing the temporal dimension into embedding and aggregation and dynamically fusing multimodal representations through gating weights \cite{liu2025multimodal}, and ontology-guided methods achieve concept-level explanation in the music domain \cite{rumiantcev2026ontology}.

\section{Evaluation and Resources}

\subsection{Explainability Evaluation Metrics}

\input{table_metrics_explainbility}

\textit{Model Transparency/Importance Metrics.}
Metrics in this category quantify the degree to which explanations align with the model's actual decision mechanism. Path Score scores each mined multi-hop user--item path in the knowledge graph by summing the products of entity and relation embeddings along the path, where higher scores indicate stronger traceability of the recommendation result \cite{zhang2025hkgat}. Integrated Gradients attributes each input feature by integrating the model's output gradients along the path from a baseline to the actual input, quantifying each feature's contribution to the prediction across modalities \cite{rychalska2020know}. DeepSHAP estimates feature attribution in deep networks through Shapley values, computing the average marginal contribution of each feature across feature subsets to capture the direction and magnitude of multimodal influences \cite{lee2026integrating}. The Stick-Breaking Mechanism is a nonparametric Bayesian approach that progressively allocates fractions of a unit ``stick'' to each modality, yielding per-decision modality contribution ratios that sum to one \cite{liu2026unveiling}. Uncertainty Quantification requires the model to report its prediction confidence along the reasoning chain and delegates low-confidence decisions to external sub-models according to a threshold, rendering the recommendation decision process scrutable \cite{zhang2026reasonrec}. Ontology Compatibility Score decomposes the recommendation score into embedding similarity and ontological semantic compatibility, the latter measuring the semantic consistency between recommended items and the domain ontology's concept hierarchy \cite{rumiantcev2026ontology}.

\textit{Counterfactual-Specific Explanation Metrics.}
Metrics in this category evaluate, from a causal inference perspective, the effectiveness of counterfactual interventions and the quality of the resulting explanations. Validity measures the fraction of generated counterfactuals that successfully move the target item out of the top-$K$ recommendation list, quantifying the success rate of the intervention \cite{liu2026rice}. Explanation Fidelity measures the fraction of recommended items for which a counterfactual explanation is successfully generated, namely that at least one attribute aspect differs before and after the image perturbation; Explanation Number counts the features required to form such an explanation over the items for which an explanation is successfully generated, with fewer features indicating a more concise explanation \cite{jain2024counterfactual}. ATE quantifies, at the population level, the average causal effect of removing a modality on the decision, whereas ITE quantifies the same effect for a single user--item decision, with ATE being the population expectation of ITE \cite{li2025harnessing}. CTR$_{cf}$ denotes the counterfactual click-through rate obtained by constraining an item to the baseline position $p_0$ through the $do(P=p_0)$ operator in the structural causal model, thereby blocking the spurious position-to-click path to recover the unbiased click probability; the accompanying MAE quantifies the mean absolute error of the counterfactual prediction after the intervention, with lower values indicating better debiasing performance \cite{zhou2026explainable}.

\textit{Generative Explanation Metrics.}
Metrics in this category evaluate the quality and diversity of automatically generated textual explanations. Feature Matching Ratio (FMR) measures, for each user--item pair, the proportion of the user's ground-truth preference features that appear in the generated explanation, whereas Feature Coverage Ratio (FCR) is the number of distinct features appearing in the generated explanations divided by the size of the dataset's feature vocabulary; the former quantifies how accurately an explanation matches user preferences and the latter the corpus-level diversity of the generated features \cite{geng2022improving}. Factual Score is an LLM-as-a-judge metric, adapted as a simplified version of FactScore, that uses single-step prompting to have an LLM rate the factual accuracy of generated text on a five-point scale against reference reviews, with low scores flagging hallucinated statements unsupported by those references \cite{yamanishi2025tourmllm}. BLEU-4 combines clipped n-gram precisions up to order four with a brevity penalty, emphasizing the fluency and accuracy of the generated text, whereas ROUGE-L computes an F-measure based on the longest common subsequence, balancing precision and recall to reflect the completeness of reference coverage \cite{zhang2026reasonrec}.

\textit{Human Evaluation Metrics.}
 Metrics in this category assess the subjective quality of explanations and their influence on users through questionnaires and controlled user studies. Persuasiveness gauges whether an explanation can convince users to accept a recommendation, typically measured by the recommendation acceptance rate \cite{zhou2025multimodal,guo2023towards}. Usefulness captures the practical assistance an explanation provides for users' decision-making, as rated by users in questionnaires \cite{zhou2025multimodal}. Effectiveness measures whether an explanation helps users make better decisions, commonly quantified by the change in item ratings before and after the explanation is provided \cite{guo2023towards}. Scrutability evaluates whether an explanation enables users to understand how their preferences are used and to detect and correct recommendations they judge wrong \cite{guo2023towards}. Trust measures whether an explanation increases users' confidence in the recommender system \cite{guo2023towards}. Plausibility assesses the logical soundness of an explanation, namely whether it is consistent with users' common sense and domain knowledge and makes sense to them \cite{guo2023towards}. User Explanation Acceptance Rate records the proportion of participants who endorse an explanation as a valid justification for the recommendation \cite{zhou2026explainable}.

\subsection{Review of Datasets}

Dataset selection determines which recommendation behaviors and explanation outcomes a study can evaluate. Table~\ref{tab:datasets} organizes the datasets used in the reviewed studies by application domain, reporting their modalities and scales, identifying the studies that use each dataset, and providing links to publicly available source datasets. This overview helps researchers select suitable evaluation resources and compare dataset characteristics across studies.

\textit{E-commerce.}
E-commerce datasets provide user interaction records and product content for studying user preferences and generating recommendation explanations. \textbf{Amazon Product Review Datasets} contain user reviews, ratings, product metadata, and images. Category-specific subsets have been used to study personalized visual explanations \cite{chen2019personalized}, joint visual and textual explanations \cite{wu2020visual}, and natural-language recommendation explanation generation \cite{ma2025explainrec}. The 2023 release contains 571.54 million reviews \cite{hou2026bridging}. \textbf{ESCI} provides shopping queries and query--product relevance annotations with four labels: Exact, Substitute, Complement, and Irrelevant \cite{reddy2022shopping}. These annotations support product relevance modeling based on users' expressed information needs. \textbf{Amazon-C4} focuses on product retrieval from complex contextual queries and contains 21,223 examples constructed by prompting ChatGPT to rewrite selected five-star Amazon reviews as first-person queries \cite{hou2026bridging}. Lin and He \cite{lin2026reinforcement} describe augmenting ESCI and Amazon-C4 with product images for multimodal recommendation explanation generation. \textbf{Pixel-1M}, used in ReasonRec \cite{zhang2026reasonrec}, contains over one million users, approximately 100,000 images, and 20 million user--image interactions, supporting preference modeling from visual content and interaction histories.

\textit{Fashion.} 
Fashion datasets support outfit compatibility modeling, personalized outfit generation, and attribute-based explanations. \textbf{Polyvore Outfits} provides clothing images, short textual descriptions, and outfit compositions for compatibility evaluation. Laenen and Moens~\cite{laenen2022learning} use both non-disjoint and disjoint versions, with the latter preventing item overlap across dataset splits. \textbf{Polyvore-U} and \textbf{iFashion} are used in DualFashion for personalized fill-in-the-blank and complete outfit generation~\cite{yu2026dual}. To represent item characteristics, the study derives structured captions from clothing images, covering color, material, design features, and style.\textbf{FashionVC} contains top--bottom combinations with item images, titles, and category labels, while \textbf{PolyvoreMaryland} provides outfit compositions. AFRec uses these datasets to extract clothing attributes and explain complementary clothing recommendations~\cite{li2021attribute}. \textbf{IQON3000} contains user-created outfits with item images, categories, and descriptions. Its association between users and outfits enables PAI-BPR~\cite{sagar2020pai} to model individual compatibility preferences and provide attribute-based interpretations.

\textit{Media content.} Media datasets combine preference signals with movie, video, or music information for recommendation and explanation studies. \textbf{MovieLens} provides user ratings and movie metadata, supporting collaborative preference modeling~\cite{harper2015movielens}. \textbf{MicroLens} connects user--video interactions with video content; MLLM-MSR uses video introductions and cover images to derive textual and visual representations of users' viewing histories~\cite{ye2025harnessing}. \textbf{YouTube-8M} provides video-level labels and precomputed visual and audio features for video understanding~\cite{abu2016youtube}, and is used by Yang et al.~\cite{yang2025research} in video recommendation research. These resources offer different forms of evidence for modeling preferences, from explicit ratings to the semantic content of consumed videos.\textbf{KKBOX} provides listening interactions and song information for music recommendation. Tao et al.~\cite{tao2021multi} use it for knowledge-aware recommendation and path-based explanations, connecting recommended songs to users through relational information. \textbf{KuaiRec} contains user feedback on short videos and offers a small subset with a nearly fully observed user--item interaction matrix~\cite{gao2022kuairec}. This subset allows recommendation evaluation with fewer missing observations. Across these datasets, ratings and behavioral feedback support preference modeling, while media content and relational information provide evidence for constructing explanations. Their different feedback structures and content representations should guide benchmark selection for the recommendation and explanation tasks under study.

\input{table_datasets}

\textit{Locations and local services.} Location-based and local service datasets connect user activities with places, reviews, and photographs, providing information about both individual preferences and destination characteristics. \textbf{Gest} contains Google Local reviews and user-uploaded images. Yan et al.~\cite{yan2023personalized} use these associations to construct tasks for personalized image-set selection and visually aware explanation generation. The combination of reviews and photographs supports the selection of images that reflect a user's interests and the generation of explanations grounded in visual and textual content. \textbf{Jalan.net Tourism Data} combines destination photographs and reviews. TourMLLM uses these resources to construct instruction data for landmark recognition, review generation, image captioning, and next-location recommendation~\cite{yamanishi2025tourmllm}. These tasks connect the visual appearance and textual descriptions of destinations with travel histories, supporting the study of multimodal tourism recommendation.

\textit{Other domains.} The reviewed studies also cover educational, conversational, and cultural heritage settings. \textbf{ASSISTments} provides student response records and knowledge-skill information, which EGRec uses for knowledge-aware educational recommendation~\cite{cen2025egrec}. \textbf{MMConv} contains multimodal, multi-turn dialogues about local venues, together with venue information, images, and annotations of dialogue states and acts. These resources support state-graph reasoning for conversational recommendation, where user preferences and recommendation requests are expressed across dialogue turns~\cite{wu2022state}. \textbf{Documerica} is a collection of documentary photographs used by Arnold and Tilton~\cite{arnold2024explainable} for explainable visual search and discovery. Their approach generates image descriptions with a multimodal language model and uses textual embeddings to retrieve photographs, extending the reviewed applications to cultural heritage exploration.

\section{Future Research Directions}

\subsection{Explanation as Evidence}

Some multimodal recommendation models have begun to leverage user history as well as various item content information such as images, text, and audio to generate natural language rationales, yet the linguistic plausibility of explanations does not guarantee their faithfulness to model decisions. Generative models may fabricate information absent from the input, or may write item attributes that objectively exist but did not actually influence the recommendation result into the explanation. Future research needs to further improve the consistency between recommendation rationales and multimodal evidence, so that generated explanations can correspond to specific historical behaviors, text attributes, image regions, audio segments, or knowledge paths.

Multimodal Retrieval-Augmented Generation (Multimodal RAG) \cite{chen2022murag} provides a direct avenue for obtaining the aforementioned evidence, as it can retrieve images and text relevant to the current task from external multimodal knowledge bases. When applied to recommendation, it can further incorporate user history and structured knowledge to supplement candidate evidence pertinent to the current decision for explanation generation. Specifically, retrieved multimodal evidence can serve as conditional input for explanation generation, grounding recommendation rationales as much as possible in traceable information. In terms of cross-modal association, vision-language models such as CLIP \cite{radford2021learning} and SigLIP \cite{zhai2023sigmoid} can establish semantic connections between image content and text concepts, forging correspondences between attribute descriptions in explanations and relevant visual evidence. CLIP leverages semantic similarity between text descriptions and image representations to evaluate whether a given attribute in an explanation is supported by visual content. By comparison, SigLIP adopts image-text matching learning to optimize cross-modal semantic alignment, rendering the matching between explanation attributes and visual evidence more stable.

\subsection{Multimodal Latent Preference Discovery}

Multimodal content such as images, text, and audio is typically encoded into high-dimensional latent representations to characterize the complex relationships between item features and user preferences. While this representational approach facilitates the capture of implicit information across different modalities, its internal factors often lack clear human-understandable semantics. Existing disentanglement methods attempt to distinguish modality-specific factors from cross-modal shared factors \cite{wang2021multimodal,laenen2022learning}, yet direct correspondences between these factors and specific preference concepts remain absent. How to identify meaningful preference concepts from the latent space and establish reliable connections between latent factors and understandable semantics remains a problem that has not been fully resolved.

One feasible approach is to introduce explicit semantic structures between latent representations and recommendation results. The concept bottleneck model \cite{koh2020concept} sets an intermediate layer composed of concepts such as color, material, style, and usage scenarios, enabling recommendation predictions to pass through interpretable concepts, thereby mapping some latent factors to specific semantic attributes. This structure also facilitates analysis of the relative roles of different intermediate concepts in final recommendation predictions. For latent spaces lacking predefined concepts, the semantic knowledge of large language models can be used to adjust clustering perspectives and granularity, identify semantically similar preference factors \cite{zhang2023clusterllm}, and assist in explaining the user interests represented by different latent clusters. Taking item content and interaction records corresponding to each latent cluster as semantic input, the model generates candidate concept labels and compares semantic consistency across different clustering granularities. Multimodal contrastive learning \cite{jia2021scaling} constructs a shared semantic space by pulling representations of matched image-text pairs closer and pushing mismatched content apart, making cross-modal correspondences another basis for discovering latent preference concepts. Shared representations of user-interacted items across different modalities can also be used to identify semantically stable factors that persist across modalities, treating them as candidates for latent preference concepts.

\subsection{Interactive Explanation}

Generative explanation provides an intrinsic pathway to multimodal interpretability by jointly deriving recommendation results and explanatory content through generative models. Existing work has explored variational autoencoders to learn disentangled representations with favorable generalizability, robustness, and interpretability, hierarchical graph generative networks for attribute-level compatibility modeling, and visualization-based methods for personalized explanation generation \cite{wang2021multimodal}. Nonetheless, current generative explanation methods still face fundamental limitations. Disentanglement quality relies heavily on weak supervision from attribute labels, yet real-world multimodal data presents extensive visual variation, noisy product descriptions, and complex cross-modal relations that complicate effective factor separation. Moreover, automatically interpretable factors remain limited to textual attributes, and spurious correlations captured during training can undermine explanation faithfulness. Future research needs to enhance the robustness and faithfulness of generative explanations in multimodal recommendation.

One promising direction is the design of self-supervised pretraining tasks specifically tailored for e-commerce multimodal data, which can yield better representations of fine-grained discriminative regions and reduce dependence on manually annotated attribute labels. Alternative reconstruction loss functions more suitable for e-commerce images can also be explored to adjust the saliency definition of variational autoencoders, better capturing domain-relevant factors of variation \cite{laenen2022learning}. For graph-structured multimodal scenarios, hierarchical generative models that organize attributes, items, and relational patterns at different levels can provide multi-granularity explanations from specific features to overall compatibility judgments \cite{guan2022multimodal}. The scope of automatically interpretable factors can further be expanded by extending cross-modal alignment mechanisms beyond textual attribute coverage.

\subsection{Evolving User Models}

Unlike relatively stable item attributes, user preferences may continuously change over time, context, and tasks. Nonetheless, some multimodal recommendation methods still aggregate historical interactions into a single or relatively fixed user representation, making it difficult to clearly distinguish long-term interests, short-term intentions, and incidental behaviors, and also limiting the description of the preference evolution process. When user representations cannot explicitly record these changes, the system similarly struggles to explain why recommendation results and their explanations shift over time. Therefore, dynamic user modeling not only needs to promptly absorb new behavioral information but also should prevent relatively stable long-term interests from being overwritten by short-term behaviors.

To address this issue, continual learning \cite{kirkpatrick2017overcoming} can update user representations as new behaviors arrive, and alleviate forgetting of existing interests by constraining important parameters or retaining historical knowledge, thereby striking a balance between new preference learning and long-term information retention. Meta-learning \cite{finn2017model} constructs related tasks using different users or different preference stages, enabling the model to quickly adjust user representations with a small number of new interactions; its core lies in learning transferable patterns in the preference change process, rather than retraining the model for each change. On the other hand, large language models \cite{salemi2024lamp} can retrieve and integrate information from user history relevant to the current task, generating personalized representations or text descriptions containing preference semantics, thereby providing richer semantic information for dynamic user modeling.

\subsection{Context-Grounded Recommendation}

Users' actual choices are not solely determined by long-term preferences and item content; their location, time, current task, and interaction process may likewise alter their decisions. Existing multimodal recommendation typically incorporates these contextual factors as auxiliary features for prediction, yet rarely further explains how the importance of each modality shifts under different contexts, and why such changes lead to different recommendation results. Consequently, future research also needs to more directly integrate contextual information into the multimodal decision-making process, so that recommendation rationales can reflect the user's current environment and interaction state.

For spatial context, vision-language models \cite{li2023blip} can jointly analyze environmental images and text descriptions, pick out locations, scenes, and objects, and give clues for judging the user's current intention. Once user historical preferences are brought in, the model can link the current environment to user needs and adjust recommendation results. For continuous interaction scenarios, embodied intelligence \cite{savva2019habitat} offers another modeling angle---linking environmental perception, action, and feedback, and updating recommendation strategies as new information arrives during interaction. This direction is about state updates and decision adjustments of recommendation systems in dynamic environments, letting the model refine the recommendation process based on user behavioral feedback. Digital twin \cite{sharma2022digital} builds virtual representations of users, items, and environments, simulates system states when time, location, or interaction conditions change, and analyzes their effects on user choices and recommendation results by tweaking relevant factors in the virtual environment.

\subsection{Hierarchical Preference Reasoning}

Existing multimodal recommendation often compresses user preferences at different levels into a unified representation space, making it difficult to determine whether a recommendation is primarily influenced by immediate perception, long-term interests, contextual trade-offs, or high-level value orientations. The research reviewed in this paper focuses more on perceptual preferences and long-term interests reflected by multimodal content and historical behavior \cite{wang2021multimodal,laenen2022learning}, while involving less strategic trade-offs under contextual constraints and high-level value orientations, and rarely explains how preferences at different levels jointly influence recommendation results.

From a modeling perspective, hierarchical structures provide a relatively natural framework for connecting user preferences of different granularities. The hierarchical Transformer encodes local multimodal features, semantic attributes, historical interests, and high-level user goals separately, and establishes connections between preferences of different granularities through inter-layer information transfer \cite{liu2019hierarchical}. Compared with compressing all preference factors into a unified representation space, hierarchical modeling can better distinguish information of different granularities and supports layer-by-layer expression from concrete content features to abstract user goals. The hierarchical graph neural network can organize image regions, text words, semantic attributes, items, usage scenarios, and user goals into different levels of nodes, describing the association between bottom-level perceptual evidence and high-level preference judgments through graph structures \cite{ying2018hierarchical}. Information propagation between nodes further describes relationships among factors at different levels, serving to analyze the role of different multimodal evidence in the recommendation process.

\subsection{Explanation-Driven Learning}

If explanations remain merely the final output of the recommendation process, users' acceptance, questioning, or modification of explanations can hardly truly influence subsequent model learning. Existing work primarily uses manually constructed explanation data to provide supervision \cite{guo2023towards}, or converts automatic evaluation results into reward signals \cite{lin2026reinforcement}, but neither approach can fully reflect real users' understanding and judgment of recommendation rationales. The absence of such feedback also means that the model struggles to timely detect erroneous attributes, irrelevant visual evidence, and inaccurate preference descriptions. Future research therefore needs to consider reintroducing users' evaluations and revisions of recommendation rationales into the learning process, for the continuous improvement of preference modeling, recommendation reasoning, and explanation generation.

This direction can unfold around different types of feedback signals. RLHF learns a reward model based on users' comparisons and evaluations of different recommendation rationales, and accordingly optimizes explanation generation strategies \cite{ouyang2022training}, transforming user feedback into optimization signals that guide explanation generation. Inverse reinforcement learning, by contrast, infers latent goals from users' behaviors of accepting, rejecting, or modifying recommendations \cite{fu2017learning}, focusing not on directly fitting the behaviors themselves but on recovering the preference basis behind these behaviors. This approach provides a reverse modeling method for inferring latent preferences using user behavioral feedback. In comparison, human-in-the-loop learning allows users to directly mark erroneous attributes, irrelevant evidence, or inaccurate preference descriptions, and applies these corrections to subsequent training \cite{wu2022survey}, enabling the model to adjust for specific explanation biases rather than relying solely on overall performance metrics for optimization.

\section{Conclusion}

Multimodal recommender systems have become a central infrastructure for online content platforms, yet the opacity of cross-modal representations and decision processes remains a persistent barrier to their trustworthy deployment. This paper offers a systematic review of interpretability techniques for multimodal recommendation. Grounded in the timing and manner of explanation generation, existing methods are organized into two categories: modality explanation and post-hoc explanation. The former covers four progressive levels---alignment, contribution, intra-modal and inter-modal---whereas the latter spans feature importance analysis, multimodal LLM-generated explanation, knowledge graph reasoning paths, and counterfactual explanation. We further synthesize four categories of interpretability evaluation metrics, and classify representative datasets along application domain and data source. Building on this taxonomy, we identify seven promising research directions, including explanation as evidence, multimodal latent preference discovery, generative explanation, and others. We hope this survey provides a unified analytical perspective across multimodal representation, recommendation decision-making, and user preferences, laying a foundation for transparent and trustworthy multimodal recommender systems.

\bibliographystyle{IEEEtran}
\bibliography{references}

\end{document}

%% file: table_modality_explainability.tex
\begin{table*}[p]
\centering
\begingroup
\footnotesize
\setlength{\tabcolsep}{2.5pt}
\renewcommand{\arraystretch}{1.10}

\caption{Summary of Modality-Level Explainability Methods}
\label{tab:modality_explainability}

\begin{tabularx}{\textwidth}{@{}P{2.0cm}P{2.8cm}P{2.8cm}P{1.3cm}YY@{}}
\toprule
Explanation category & Study & Main method & Content modalities
& Evaluation metric & Dataset \\
\midrule
\multirow{8}{2.0cm}{\centering Modality Alignment Explanations} & EdgeX-MMFRec \cite{alamdari2025edgex} & Attribution Learning & I, T, A & HR; NDCG; MRR; Attribution-Faithfulness; BLEU; ROUGE; Human evaluation; etc. & Amazon-MM; Yelp-Photos; MovieLens-MM; Last.fm-MM  \\
& EARec \cite{yang2025explainable} & Large Language Model & V, T & HR; NDCG; ROUGE-L & Amazon-Food; Amazon-CDs; Amazon-Kindle; Amazon-Movies; Amazon-Home; Amazon-Office; Amazon-Arts; Amazon-Instruments; MovieLens \\
& VTJEI \cite{wu2020visual} & Attention Mechanism & V, T & HR; NDCG; F1 & Amazon-Men; Amazon-Women; Amazon-Boys\&Girls; Amazon-Baby \\
& MDR \cite{wang2021multimodal} & Disentangled Representation Learning & V, T & AUC; NDCG & Amazon-Clothing; Amazon-Toys; Amazon-Office \\
& MMALFM \cite{cheng2019mmalfm} & Topic Modeling & V, T & RMSE; MAE; Precision; HR; NDCG & Amazon-Beauty; Amazon-CDs; Amazon-Clothing; Amazon-Movies; Amazon-Cell Phones \\
& Laenen and Moens \cite{laenen2022learning} & Disentangled Representation Learning & V, T & AUC; Accuracy; Recall & Polyvore Outfits; Amazon-Dresses \\
& DGVAE \cite{zhou2024disentangled} & Disentangled Representation Learning & V, T & Recall; NDCG & Amazon-Baby; Amazon-Sports; Amazon-Clothing \\
& METER \cite{geng2022improving} & Contrastive Learning & V, T & FMR; FCR; USR; BLEU-4; CLIPScore; RMSE; MAE; etc. & TripAdvisor; Yelp \\
\midrule
\multirow{6}{2.0cm}{\centering Modality Contribution Explanations} & MMT \cite{zhao2025multimodal} & Attention Mechanism & V, T, A & Precision; Recall; NDCG; AUC; F1 & Short-video Proprietary Dataset; KUBD \\
& GraphFusion-HRL \cite{zare2025graphfusion} & Attention Mechanism & V, T & Recall; NDCG; Diversity; Coverage; MEG & MovieLens-1M; Amazon-Fashion; Amazon-Books; ModCloth; Amazon-Mixed \\
& Yang et al. \cite{yang2025research} & Attention Mechanism & V, T, A & Precision; Recall; Privacy Loss; Latency & YouTube-8M \\
& AM-Bi-LSTM \cite{ohtomo2024bi} & Attention Mechanism & V, T & AUC; MRR; Recall; CC & Forestanet; Enhanced MovieLens-100K \\
& MDLA \cite{liu2026unveiling} & Attention Mechanism & V, T & F1; NDCG; ROUGE & Amazon-Clothing; Amazon-Shoes; Amazon-CPA \\
& MTR \cite{liu2025multimodal} & Mixture of Experts & V, T & MRR; Hit; Recall & NYC; Yelp; Amazon-Electronics \\
\midrule
\multirow{8}{2.0cm}{\centering Intra-Modality Explanations} & MLLM-MSR \cite{ye2025harnessing} & Multimodal Large Language Model & V, T & AUC; HR; MRR & Microlens; Amazon-Baby; Amazon-Game \\
& AFRec \cite{li2021attribute} & Attention Mechanism & V, T & AUC; HR & FashionVC; PolyvoreMaryland \\
& MUA-RS \cite{torres2023multimodal} & User-adaptive Filtering & V & Precision; Recall; NDCG & MovieLens-100K \\
& PS-OCM \cite{guan2022multimodal} & Disentangled Representation Learning & V, T & AUC; MRR & IQON3000 \\
& VECF \cite{chen2019personalized} & Attention Mechanism & V, T & F1; HR; NDCG & Amazon-Men; Amazon-Women; Amazon-Boys\&Girls; Amazon-Baby \\
& Re-VECF \cite{chen2018visually} & Attention Mechanism & V, T & F1; NDCG & Amazon-Men; Amazon-Women \\
& Ante-RNN \cite{liu2020dynamic} & Attention Mechanism & V, T & Recall; NDCG & MovieLens; Pinterest \\
& LIP \cite{jian2026large} & Intent Disentanglement & V, T & Recall; NDCG & Amazon-Baby; Amazon-Beauty; Amazon-Toys \\
\midrule
\multirow{7}{2.0cm}{\centering Inter-Modality Explanations} & MKRLN \cite{tao2021multi} & Knowledge Graph & V, T & NDCG; Precision;  & Book; Movie; KKBOX \\
& SGR \cite{wu2022state} & Knowledge Graph & V, T & Act Accuracy; EMR; IMR; SR & MMConv \\
& CDPRec \cite{sang2020context} & Attention Mechanism & V, T & Prec; Recall; F-score; AUC & YouTube video dataset \\
& PC2L \cite{yan2023personalized} & Contrastive Learning & V, T & Precision; Recall; F1; BLEU; METEOR; NIST; CLIP-Align; BERT-Score & Gest (Google Local Restaurants) \\
& MMCT \cite{liu2023multimodal} & Contrastive Learning & V, T & RMSE; MAE; NDCG; HR; BLEU; ROUGE; USR; FMR; FCR; DIV & Amazon-Movies; Amazon-Clothing \\
& MRG \cite{truong2019multimodal} & Attention Mechanism & V, T & BLEU; ROUGE-L; MAE; RMSE & Yelp (Chicago, Los Angeles, New York, San Francisco) \\
& Rumiantcev et al. \cite{rumiantcev2026ontology} & Knowledge Graph & A, T & Precision; Recall; F1; MAP & Million Song Dataset; GTZAN; FSD50K; FMA \\
\bottomrule
\end{tabularx}
\endgroup
\end{table*}

%% file: table_posthoc_explainability.tex
\begin{table*}[!t]
\centering
\begingroup
\footnotesize
\setlength{\tabcolsep}{2.5pt}
\renewcommand{\arraystretch}{1.10}

\caption{Summary of Post-Hoc Explainability Methods}
\label{tab:posthoc_explainability}

\begin{tabularx}{\textwidth}{@{}P{2.0cm}P{2.8cm}P{2.8cm}P{1.3cm}YY@{}}
\toprule
Explanation category & Study & Main method & Content modalities
& Evaluation metric & Dataset \\
\midrule

\multirow{2}{2.0cm}{\centering Feature Importance Analysis}
& Rychalska et al.~\cite{rychalska2020know}
& Attribution Learning
& V, T
& Attribution score
& MovieLens-20M \\

& FusionNet~\cite{lee2026integrating}
& SHAP
& V, T
& MAE; SHAP Value
& Amazon-Automotive; Amazon-Digital Music; Amazon-Musical Instruments; Amazon-Office Products \\

\midrule

\multirow{11}{2.0cm}{\centering Multimodal LLM-Generated Explanations}
& HCR~\cite{liang2025enhancing}
& Prompt Tuning
& V, T, A
& HR; NDCG; USR
& Amazon-Arts; Amazon-Fashion; Amazon-Software; Amazon-Game; Healthcare-Video; Healthcare-Doctor \\

& MLLM-MSR~\cite{ye2025harnessing}
& Preference Summarization
& V, T
& AUC; HR; MRR
& MovieLens; Amazon-Baby; Amazon-Game \\

& TourMLLM~\cite{yamanishi2025tourmllm}
& Retrieval-Augmented Generation
& V, T
& Accuracy; BLEU; CIDEr; BERT-F1; CLIP-BERT; etc.
& Jalan.net Tourism Data \\

& MSRBench~\cite{zhou2025large}
& Reranking
& V, T
& HR; NDCG
& Amazon-Beauty; Amazon-Sports; Amazon-Toys; Amazon-Clothing \\

& MMP-Refer~\cite{pan2026mmp}
& Multimodal Path Retrieval
& V, T
& GPTScore; BERTScore; BARTScore; BLEURT; USR
& Amazon-Baby; Amazon-Sports; Amazon-Clothing \\

& ExplainRec~\cite{ma2025explainrec}
& Preference attribution
& V, T
& AUC; Hit; NDCG; BLEU
& MovieLens-25M; Amazon-Movies \& TV; Amazon-Books  \\

& DynMM-Explain-LLMRec~\cite{ma2025bridging}
& Evidence-grounded Generation
& V, T, A
& Hit; NDCG; Recall; BLEU; Faithfulness; Consistency
& Amazon-Movies \& TV; Amazon-Video Games; Amazon-Beauty; Amazon-Toys; KuaiRec; MovieLens-1M  \\

& MM-Rec-RL~\cite{lin2026reinforcement}
& Reinforcement Learning
& V, T
& Recall; NDCG; BLEU; Simulated User Satisfaction
& Amazon-Beauty; ESCI; Amazon-C4  \\

& ReasonRec~\cite{zhang2026reasonrec}
& Chain-of-Thought Reasoning
& V, T
& HR; NDCG; BLEU-4; ROUGE-L; AUC; LogLoss
& Amazon-Sports; Amazon-Beauty; Amazon-Clothing; Amazon-Toys; Pixel-1M \\

& ENF~\cite{chen2026top}
& Multi-Agent Reasoning
& V, T
& Accuracy; Precision; Recall; F1; Class Accuracy; Reasoning Score
& TVNF \\

& Arnold and Tilton~\cite{arnold2024explainable}
& Caption-based Explanation
& V, T
& Term matching accuracy
& Documerica  \\

\midrule

\multirow{3}{2.0cm}{\centering Knowledge Graph-Based Reasoning Paths}
& HKGAT~\cite{zhang2025hkgat}
& Path Analysis
& V, T
& Precision; Recall; F1; NDCG
& GAD; Amazon-Book \\

& EGRec~\cite{cen2025egrec}
& Semantic Path Explanation
& V, T
& HR; NDCG; MRR
& ASSISTments; XueTangX \\



& PAI-BPR~\cite{sagar2020pai}
& Attribute Importance
& V, T
& AUC; MRR
& IQON3000 \\

\midrule

\multirow{4}{2.0cm}{\centering Counterfactual explanation}
& RICE~\cite{liu2026rice}
& Vector Search
& V, T
& Validity; Diversity; Distance
& Amazon-Baby; Amazon-Sports; Amazon-Clothing \\

& CAVIAR~\cite{jain2024counterfactual}
& Embedding Perturbation
& V, T
& Fidelity; Explanation Fidelity; Explanation Number
& Amazon-Fashion-Men; Amazon-Fashion-Women \\

& MACE~\cite{zhou2025multimodal}
& Comparative Reasoning
& V, T
& Persuasiveness; Helpfulness; Specificity; Clarity
& Amazon-Clothing \\

& Zhou et al.~\cite{zhou2026explainable}
& Dual-path Reasoning
& V, T
& NDCG; Long-tail Coverage; User Interpretation Acceptance Rate; etc.
& MovieLens-20M \\

\bottomrule
\end{tabularx}
\endgroup
\end{table*}

%% file: table_metrics_explainbility.tex
\begin{table*}[!t]
\caption{Explainability Metrics Used in the Reviewed Multimodal Explainable Recommendation Studies}
\label{tab:metrics_explainability}
\centering
\scriptsize
\setlength{\tabcolsep}{1.8pt}
\renewcommand{\arraystretch}{0.95}
\begin{tabularx}{\textwidth}{P{2.5cm}P{3.0cm}YP{1.55cm}}
\toprule
Category & Metric & Definition & Studies in this survey \\
\midrule
\multirow{6}{2.5cm}{\centering Model Transparency/Importance Metrics}
& Path Score
& Quantifies user--item multi-hop reasoning paths in the knowledge graph via relation-type weights and hop distance.
& \cite{zhang2025hkgat}, \cite{cen2025egrec}, \cite{pan2026mmp} \\
& Integrated Gradients
& Attributes each input feature by integrating output gradients along the path from a baseline to the actual input.
& \cite{rychalska2020know} \\
& DeepSHAP
& Estimates each feature's average marginal contribution via Shapley values, capturing the direction and magnitude of multimodal influences.
& \cite{lee2026integrating} \\
& Stick-Breaking Mechanism
& Progressively allocates proportions of a unit ``stick'' to each modality, yielding per-decision contribution ratios summing to one.
& \cite{liu2026unveiling} \\
& Uncertainty Quantification
& Requires the model to output prediction confidence, quantifying epistemic and aleatoric uncertainty to render decision boundaries scrutable.
& \cite{zhang2026reasonrec} \\
& Ontology Compatibility Score
& Decomposes the score into embedding similarity and ontological semantic compatibility with the domain concept hierarchy.
& \cite{rumiantcev2026ontology} \\
\midrule
\multirow{7}{2.5cm}{\centering Counterfactual-Specific Explanation Metrics}
& Validity
& Proportion of counterfactual scenarios that successfully flip the recommendation result.
& \cite{liu2026rice} \\
& Explanation Fidelity
& Proportion of successfully generated counterfactual explanations that remain readable and semantically coherent.
& \cite{jain2024counterfactual} \\
& Explanation Number
& Number of features that must be modified to flip the recommendation result.
& \cite{jain2024counterfactual} \\
& ATE
& Average causal effect of removing a modality on recommendation decisions at the population level.
& \cite{li2025harnessing} \\
& ITE
& Causal effect of removing a modality on a single user--item decision, of which ATE is the population expectation.
& \cite{li2025harnessing} \\
& CTR\_cf
& Virtual click-through rate computed after counterfactual intervention on bias nodes via the do-operator.
& \cite{zhou2026explainable} \\
& MAE
& Prediction error of bias nodes after counterfactual intervention; lower values indicate better debiasing.
& \cite{zhou2026explainable} \\
\midrule
\multirow{5}{2.5cm}{\centering Generative Explanation Metrics}
& FMR
& Proportion of the user's genuine preference features contained in the generated explanation.
& \cite{geng2022improving} \\
& FCR
& Proportion of item feature dimensions covered by the generated explanation.
& \cite{geng2022improving} \\
& Factual Score
& Consistency between generated explanations and ground-truth factual descriptions, used to detect hallucination.
& \cite{yamanishi2025tourmllm} \\
& BLEU-4
& Evaluates generated explanations through 4-gram matching precision with a length penalty.
& \cite{zhang2026reasonrec} \\
& ROUGE-L
& Evaluates coverage of reference explanations through the longest common subsequence.
& \cite{zhang2026reasonrec} \\
\midrule
\multirow{7}{2.5cm}{\centering Human Evaluation Metrics}
& Persuasiveness
& Ability of explanations to persuade users to accept the recommendation.
& \cite{zhou2025multimodal}, \cite{guo2023towards} \\
& Usefulness
& Practical help of explanations for users' decision-making.
& \cite{zhou2025multimodal}, \cite{guo2023towards} \\
& Effectiveness
& Whether explanations successfully convey the recommendation rationale so that users understand the decision basis.
& \cite{guo2023towards} \\
& Scrutability
& Whether users can examine and challenge recommendation decisions and adjust the results accordingly.
& \cite{guo2023towards} \\
& Trust
& Effect of explanations on boosting users' overall trust in the recommender system.
& \cite{guo2023towards} \\
& Plausibility
& Consistency of explanations with the model's true rationale and users' common sense and domain knowledge.
& \cite{guo2023towards} \\
& User Explanation Acceptance Rate
& Proportion of participants who endorse an explanation as a valid justification.
& \cite{zhou2026explainable} \\
\bottomrule
\end{tabularx}
\end{table*}

%% file: table_datasets.tex
\begin{table*}[!t]
\caption{Datasets Used in the Reviewed Multimodal Explainable Recommendation Studies}
\label{tab:datasets}
\centering
\scriptsize
\setlength{\tabcolsep}{1.8pt}
\renewcommand{\arraystretch}{0.95}
\begin{tabularx}{\textwidth}{>{\centering\arraybackslash}p{2.5cm}P{3.5cm}P{1.5cm}P{6.20cm}Y}
\toprule
Domain & Dataset & Modalities & Scale & Studies in this survey \\
\midrule
\multirow{4}{=}{\centering E-commerce}
& Amazon Product Review Datasets (2013--2023)\footnotemark[1] & V, T & 6.64M--54.51M users; 2.44M--48.19M items; 34.69M--571.54M reviews & \cite{alamdari2025edgex,wu2020visual,zare2025graphfusion,chen2019personalized,chen2018visually,ye2025harnessing,liu2025multimodal,liang2025enhancing,zhou2025large,pan2026mmp,ma2025explainrec,ma2025bridging,lin2026reinforcement,zhang2026reasonrec,zhang2025hkgat,liu2026rice,jain2024counterfactual,zhou2025multimodal,lee2026integrating,yang2025explainable,wang2021multimodal,cheng2019mmalfm,zhou2024disentangled} \\

& ESCI \footnotemark[2] & T & 130K queries; 2.6M query-product pairs & \cite{lin2026reinforcement} \\
& Amazon-C4 \footnotemark[3] & T & 21,223 queries & \cite{lin2026reinforcement} \\
& Pixel-1M\footnotemark[4] & V & 1M+ users; 100K images; 20M interactions & \cite{zhang2026reasonrec} \\
\midrule
\multirow{7}{=}{\centering Fashion}
& Polyvore Outfits\footnotemark[5] & V, T & 68K outfits; 261K items & \cite{laenen2022learning} \\

& Polyvore-U\footnotemark[6] & V, T &  & \cite{yu2026dual} \\
& iFashion\footnotemark[6] & V, T & 344K items; 50 categories & \cite{yu2026dual} \\
& FashionVC\footnotemark[7] & V, T & 20.7K outfits; 14.9K tops; 13.7K bottoms & \cite{li2021attribute}\\
& PolyvoreMaryland\footnotemark[8] & V, T & 21.9K outfits & \cite{li2021attribute}  \\

& IQON3000\footnotemark[9] & V, T & 217.8K outfits; 3,568 users & \cite{sagar2020pai} \\
& ModCloth\footnotemark[10]& V, T & 8K users; 3K items; 0.5M interactions & \cite{zare2025graphfusion} \\
\midrule
\multirow{5}{=}{\centering Media content}
& MovieLens (100K, 1M, 20M, and 25M)\footnotemark[11] & T & 0.94K--162.5K users; 1.68K--62.4K movies; 100K--25M ratings & \cite{alamdari2025edgex,zare2025graphfusion,liu2020dynamic,rychalska2020know,ma2025explainrec,ma2025bridging,zhou2026explainable,yang2025explainable,ohtomo2024bi} \\
& Microlens\footnotemark[12] & V, T & 25K users; 20K videos; 223K interactions & \cite{ye2025harnessing} \\
& YouTube-8M\footnotemark[13] & V, T & 6.1M videos; 3,862 classes & \cite{yang2025research} \\
& KKBOX\footnotemark[14]  & T & 2K users; 34.6K songs; 61.5K interactions & \cite{tao2021multi} \\
& KuaiRec\footnotemark[15] & T & 7.2K users; 10.7K videos; 12.5M interactions & \cite{ma2025bridging} \\

\midrule
\multirow{2}{=}{\centering Locations and local services}
& Gest\footnotemark[16] & V, T & 1.01M users; 65K businesses; 1.77M reviews; 4.44M images & \cite{yan2023personalized} \\
& Jalan.net Tourism Data\footnotemark[17] & V, T & 508.8K images; 583.2K reviews; 35.7K locations & \cite{yamanishi2025tourmllm} \\
\midrule
\multirow{3}{=}{\centering Other domains}
& ASSISTments\footnotemark[18] & T & 24K students; 1.03M records; 100+ knowledge points & \cite{cen2025egrec} \\
& MMConv\footnotemark[19]  & V, T & 5K+ dialogues; 1.7K venues; 100K+ images & \cite{wu2022state} \\
& Documerica\footnotemark[20] & V & 16K+ images & \cite{arnold2024explainable} \\

\bottomrule
\end{tabularx}
\end{table*}
\footnotetext[1]{\fontsize{6}{7}\selectfont\url{https://amazon-reviews-2023.github.io}}
\footnotetext[2]{\fontsize{6}{7}\selectfont\url{https://github.com/amazon-science/esci-data}}
\footnotetext[3]{\fontsize{6}{7}\selectfont\url{https://huggingface.co/datasets/McAuley-Lab/Amazon-C4}}
\footnotetext[4]{\fontsize{6}{7}\selectfont\url{https://github.com/westlake-repl/PixelRec}}
\footnotetext[5]{\fontsize{6}{7}\selectfont\url{https://huggingface.co/datasets/mvasil/polyvore-outfits}}
\footnotetext[6]{\fontsize{6}{7}\selectfont\url{https://drive.google.com/drive/folders/1SipoOyu0FKwxceItxAfC_O7BeWS5xwVd}}
\footnotetext[7]{\fontsize{6}{7}\selectfont\url{https://drive.google.com/file/d/1d72E3p4w280-vdCKfXtXZLMULpIScRRR/view}}
\footnotetext[8]{\fontsize{6}{7}\selectfont\url{https://drive.google.com/drive/folders/0B4Eo9mft9jwoVDNEWlhEbUNUSE0}}
\footnotetext[9]{\fontsize{6}{7}\selectfont\url{https://drive.google.com/file/d/1sTfUoNPid9zG_MgV--lWZTBP1XZpmcK8/view}}
\footnotetext[10]{\fontsize{6}{7}\selectfont\url{https://cseweb.ucsd.edu/~jmcauley/datasets.html}}
\footnotetext[11]{\fontsize{6}{7}\selectfont\url{https://grouplens.org/datasets/movielens}}
\footnotetext[12]{\fontsize{6}{7}\selectfont\url{https://github.com/westlake-repl/MicroLens}}
\footnotetext[13]{\fontsize{6}{7}\selectfont\url{https://research.google.com/youtube8m}}
\footnotetext[14]{\fontsize{6}{7}\selectfont\url{https://www.kaggle.com/c/kkbox-music-recommendation-challenge/data}}
\footnotetext[15]{\fontsize{6}{7}\selectfont\url{https://github.com/chongminggao/KuaiRec}}
\footnotetext[16]{\fontsize{6}{7}\selectfont\url{https://github.com/zzxslp/Gest}}
\footnotetext[17]{\fontsize{6}{7}\selectfont\url{https://drive.google.com/drive/folders/1Wg9JPV9pdv_eqJOR0MuSO7CGJd40UBGc}}
\footnotetext[18]{\fontsize{6}{7}\selectfont\url{https://sites.google.com/site/assistmentsdata}}
\footnotetext[19]{\fontsize{6}{7}\selectfont\url{https://github.com/liziliao/MMConv/tree/main/dataset}}
\footnotetext[20]{\fontsize{6}{7}\selectfont\url{https://catalog.archives.gov/id/542493}}